\documentclass[10pt,a4paper]{amsart}
\usepackage[margin=20mm]{geometry}
\usepackage[numbers]{natbib}
\usepackage[colorlinks]{hyperref}

\usepackage{amssymb,amsmath}
\usepackage{amsthm}
\usepackage{bbm}
\usepackage{stmaryrd}
\usepackage[usenames,dvipsnames]{xcolor}
\usepackage{color}

\input xy
\xyoption{all} 

\numberwithin{equation}{section}

\newtheorem{theorem}{Theorem}[section]

\newtheorem{proposition}[theorem]{Proposition}

\theoremstyle{definition}
\newtheorem{definition}[theorem]{Definition}
\newtheorem{remark}[theorem]{Remark}
\newtheorem{example}[theorem]{Example}

\newcommand{\Id}{\mathbbmss{1}}

\newcommand{\rmi}{ \textnormal{i}}
\newcommand{\rmd}{\textnormal{d}}

\newcommand{\rmh}{\textnormal{h}}

\DeclareMathOperator{\Vect}{Vect}

\DeclareMathOperator{\Span}{Span}

\font\black=cmbx10 \font\sblack=cmbx7 \font\ssblack=cmbx5 \font\blackital=cmmib10  \skewchar\blackital='177
\font\sblackital=cmmib7 \skewchar\sblackital='177 \font\ssblackital=cmmib5 \skewchar\ssblackital='177
\font\sanss=cmss10 \font\ssanss=cmss8 
\font\sssanss=cmss8 scaled 600 \font\blackboard=msbm10 \font\sblackboard=msbm7 \font\ssblackboard=msbm5
\font\caligr=eusm10 \font\scaligr=eusm7 \font\sscaligr=eusm5  \font\fraktur=eufm10
\font\sfraktur=eufm7 \font\ssfraktur=eufm5 
\font\bsymb=cmsy10 scaled\magstep2
\def\all#1{\setbox0=\hbox{\lower1.5pt\hbox{\bsymb
       \char"38}}\setbox1=\hbox{$_{#1}$} \box0\lower2pt\box1\;}
\def\exi#1{\setbox0=\hbox{\lower1.5pt\hbox{\bsymb \char"39}}
       \setbox1=\hbox{$_{#1}$} \box0\lower2pt\box1\;}

\def\tx#1{{\fam0\relax#1}}

\newfam\bifam
\textfont\bifam=\blackital \scriptfont\bifam=\sblackital \scriptscriptfont\bifam=\ssblackital

\newfam\blfam
\textfont\blfam=\black \scriptfont\blfam=\sblack \scriptscriptfont\blfam=\ssblack

\newfam\bbfam
\textfont\bbfam=\blackboard \scriptfont\bbfam=\sblackboard \scriptscriptfont\bbfam=\ssblackboard

\newfam\ssfam
\textfont\ssfam=\sanss \scriptfont\ssfam=\ssanss \scriptscriptfont\ssfam=\sssanss
\def\sss#1{{\fam\ssfam\relax#1}}

\newfam\clfam
\textfont\clfam=\caligr \scriptfont\clfam=\scaligr \scriptscriptfont\clfam=\sscaligr

\newfam\frfam
\textfont\frfam=\fraktur \scriptfont\frfam=\sfraktur \scriptscriptfont\frfam=\ssfraktur

\def\hpb#1{\setbox0=\hbox{${#1}$}
    \copy0 \kern-\wd0 \kern.2pt \box0}
\def\vpb#1{\setbox0=\hbox{${#1}$}
    \copy0 \kern-\wd0 \raise.08pt \box0}

\def\pmb#1{\setbox0\hbox{${#1}$} \copy0 \kern-\wd0 \kern.2pt \box0}
\def\pmbb#1{\setbox0\hbox{${#1}$} \copy0 \kern-\wd0
      \kern.2pt \copy0 \kern-\wd0 \kern.2pt \box0}
\def\pmbbb#1{\setbox0\hbox{${#1}$} \copy0 \kern-\wd0
      \kern.2pt \copy0 \kern-\wd0 \kern.2pt
    \copy0 \kern-\wd0 \kern.2pt \box0}
\def\pmxb#1{\setbox0\hbox{${#1}$} \copy0 \kern-\wd0
      \kern.2pt \copy0 \kern-\wd0 \kern.2pt
      \copy0 \kern-\wd0 \kern.2pt \copy0 \kern-\wd0 \kern.2pt \box0}
\def\pmxbb#1{\setbox0\hbox{${#1}$} \copy0 \kern-\wd0 \kern.2pt
      \copy0 \kern-\wd0 \kern.2pt
      \copy0 \kern-\wd0 \kern.2pt \copy0 \kern-\wd0 \kern.2pt
      \copy0 \kern-\wd0 \kern.2pt \box0}

\mathchardef\za="710B  
\mathchardef\zb="710C  
\mathchardef\zg="710D  
\mathchardef\zd="710E  
\mathchardef\zve="710F 
\mathchardef\zz="7110  
\mathchardef\zh="7111  
\mathchardef\zvy="7112 
\mathchardef\zi="7113  
\mathchardef\zk="7114  
\mathchardef\zl="7115  
\mathchardef\zm="7116  
\mathchardef\zn="7117  
\mathchardef\zx="7118  
\mathchardef\zp="7119  
\mathchardef\zr="711A  
\mathchardef\zs="711B  
\mathchardef\zt="711C  
\mathchardef\zu="711D  
\mathchardef\zvf="711E 
\mathchardef\zq="711F  
\mathchardef\zc="7120  
\mathchardef\zw="7121  
\mathchardef\ze="7122  
\mathchardef\zy="7123  
\mathchardef\zf="7124  
\mathchardef\zvr="7125 
\mathchardef\zvs="7126 
\mathchardef\zf="7127  
\mathchardef\zG="7000  
\mathchardef\zD="7001  
\mathchardef\zY="7002  
\mathchardef\zL="7003  
\mathchardef\zX="7004  
\mathchardef\zP="7005  
\mathchardef\zS="7006  
\mathchardef\zU="7007  
\mathchardef\zF="7008  
\mathchardef\zW="700A  
\mathchardef\zC="7009  

\newcommand{\be}{\begin{equation}}
\newcommand{\ee}{\end{equation}}

\newcommand{\bea}{\begin{eqnarray}}
\newcommand{\eea}{\end{eqnarray}}
\def\*{{\textstyle *}}
\newcommand{\R}{{\mathbb R}}

\newcommand{\C}{{\mathbb C}}
\newcommand{\Z}{{\mathbb Z}}

\newcommand{\s}{{\textstyle *}}

\def\Vect{\sss{Vect}}

\def\xi{\tx{i}}

\def\cM{\cal M}

\def\s*{{\scriptstyle *}}

\def\cO{\mathcal{O}}

\def\cM{\mathcal{M}}

\newcommand{\beas}{\begin{eqnarray*}}
\newcommand{\eeas}{\end{eqnarray*}}

\def\half{\frac{1}{2}}

\title{On a $\Z_2^n$-graded version of supersymmetry}

   \author{Andrew James Bruce} 
   \address{Mathematics Research Unit, University of Luxembourg, Maison du Nombre 6, avenue de la Fonte, 
L-4364 Esch-sur-Alzette}   
   \email{andrewjamesbruce@googlemail.com}
  
\date{\today}
 
\begin{document}

\begin{abstract}
We  extend the notion of super-Minkowski space-time to include $\Z_{2}^{n}$-graded (Majorana) spinor coordinates. Our choice of the grading leads to  spinor coordinates that are nilpotent but commute amongst themselves. The mathematical framework we employ is the recently developed category of  $\Z_{2}^{n}$-manifolds understood as locally ringed spaces. The formalism we present resembles $\mathcal{N}$-extended superspace (in the presence of central charges), but with some subtle differences due to the exotic nature of the grading employed.   \par
\end{abstract}

 \maketitle

\setcounter{tocdepth}{2}
 \tableofcontents

\noindent \emph{In geometric and physical applications, it always turns out that a quantity is characterized not only by its tensor order, but also by symmetry.} --  Hermann Weyl (1925)

\section{Introduction} 
 Supersymmetry was  independently discovered by three groups of authors: Gervais \& Sakita \cite{Gervais:1971}, Gol'fand \& Lichtman \cite{Golfand:1971} and  Volkov \& Akulov \cite{Volkov:1972} in the early 1970s. The first example of an interacting supersymmetric quantum field theory in four dimensions was due to Wess \& Zumino \cite{Wess:1974}.   This non-classical symmetry  is more than just a way of relating bosonic and fermionic degrees of freedom as supersymmetric field theories can have some remarkable  properties. For example, supersymmetry can lead to milder divergences and even `non-renormalisation theorems'; offers a solution to the hierarchy problem in Grand Unified Theories; removes the tachyon from the spectrum of string theories and naturally leads to a  theory of gravity when promoted to a local gauge theory. However, we stress  that no experimental evidence  that Nature utilises  supersymmetry has yet been found.  More than this, current experimental data suggest that the (constrained) minimal supersymmetric standard model is disfavoured as a potential realistic model (see Autermann \cite{Autermann:2016}). It is even said that ``supersymmetry is a solution looking for a problem'' (see for example Kaku \cite[page 101]{Kaku:1999}). \par
Alongside the vast wealth of results from theoretical physics, there has been a steady growth in the mathematical aspects of supersymmetry since its conception. This knowledge accumulates to the theory of supermanifolds and super Lie groups, with  the definition of a supermanifold being due to  Berezin \&  Le\u{\i}tes \cite{Berezin:1976}.  Much of the fundamental work was carried out between 1965 and 1975 by Berezin and his collaborators.  At its `bare bones' supermathematics is the study of $\Z_{2}$-graded structures. The study of such things is well motivated by physics via the fact that quasi-classical descriptions of fermions and  Faddeev--Popov ghosts both  require anticommuting fields.  From a pure mathematical perspective, various `sign rules' appear naturally in algebraic topology and homological algebra (see Mac~Lane \cite{MacLane:1963}).  \par 
Some very powerful methods for building supersymmetric actions come under the umbrella of \emph{superspace methods}, as first developed (operationally)  by Salam \& Strathdee   \cite{Salam:1974}.  These methods allow for a more geometric picture of supersymmetry as well as giving practical methods of constructing theories. However, the formalism can become clumsy, and not all supersymmetric theories have a superspace formulation. For instance, it is well known that there are no off-shell formulations of $d= 4$,  $\mathcal{N}=4$ theories with a finite number of auxiliary fields. One has to resort to projective or harmonic superspaces   in order to cover this situation (see \cite{Galperin:2007}).\par 
  At the most basic level, one postulates that  standard four-dimensional Minkowski space-time is extended by appending four anticommuting Majorana spinors. Thus, $\mathcal{N}=1$ super-Minkowski space-time is the supermanifold that admits (global) coordinates
$$(x^{\mu}, \theta_{\alpha})\,.$$
The supersymmetry transformations on this supermanifold are
\begin{align}\label{eqn:N=1}
& x^{\mu} \mapsto x^{\mu} + \frac{1}{4} \epsilon_{\beta}\theta_{\alpha}(C\gamma^{\mu})^{\alpha \beta}, && \theta_{\alpha} \mapsto \theta_{\alpha}+ \epsilon_{\alpha},
\end{align}
where $\epsilon_{\alpha}$ is an anticommuting Majorana spinor-valued parameter. For a detailed introduction to supersymmetry we suggest West \cite{West:1990},  and  Wess \& Bagger \cite{Wess:1992}. For an introduction to superspace and methods of quantising supersymmetric field theories one may consult Gates et al. \cite{Gates:1983}. One can also consult the encyclopedia edited by Duplij, Siegel \& Bagger \cite{Duplij:2004} for many details pertaining to supersymmetry and related mathematics, including historical remarks in the section ``SUSY Story'' written by  the founders of supersymmetry.
\par 
Extended supersymmetries are then constructed by appending more and more anticommuting Majorana spinors. For example, $\mathcal{N}=2$ super-Minkowski space-time comes with local coordinates $(x^{\mu}, \theta_{\alpha}^{i})$, $i = 1,2$. In particular, we still have a supermanifold  and thus $\theta^{1}\theta^{2} = {-} \theta^{2}\theta^{1}$. The case for $\mathcal{N} >2$ is analogous. Moreover,  $\mathcal{N}$-extended supersymmetry transformations similar to \eqref{eqn:N=1} exist.\par 
However, \emph{mathematically} there is no reason not to consider appending  sets of  Majorana spinors that do not necessarily anticommute. Indeed,  non-anticommuting superspaces have long been studied in the physics literature, see  for example \cite{Connes:1998,Ferrara:2003,Schwarz:2003,Schwarz:1982,Seiberg:2003,Seiberg:1999}.  The inspiration for many of these works is the well-known fact that various background fields  in string theory lead to noncommutative deformations.  For example, R-R field backgrounds lead to `$\theta-\theta$' deformations and gravitino backgrounds lead to `$x-\theta$' deformations. We will consider the very special instance of Majorana spinors that are $\Z_{2}^{n}$-graded commutative. This leads to  Majorana spinors that square to zero, yet commute amongst themselves. We view this situation as a very mild form of `$\theta-\theta$' non-anticommutativity.  This should, of course, be compared with Green's notion of a parafermion (see Green \cite{Green:1953}, and also Volkov \cite{Volkov:1959} who independently introduced the concept).  Because of  how we will assign the $\Z_{2}^{n}$-grading the supersymmetry algebra `$[Q, Q] \sim P + Z$' where we allow a central charge $Z$, is not given in terms of just an anticommutator, some terms will be given in terms of a commutator,  collectively  we use $\Z_{2}^{n}$-graded commutators. This will complicate the understanding of Bogomol'nyi-Prasad-Sommerfield (BPS) states, for example. At this juncture, we should also mention the work of  Zheltukhin \cite{Zheltukhin:1987} who defined a `para-Fermionic superspace' in relation to a generalisation of the Neveu--Ramond--Schwarz superstring. As far as we know,  Zheltukhin is the first to consider a `superspace' formed by replacing standard Grassmann coordinates with a $\Z_2^2$-graded version. We must also mention the work of Vasil'iev \cite{Vasiliev:1985}, in which a $\Z_2^2$-Lie algebra structure was uncovered in supergravity with a positive cosmological constant by modifying spinor conjugation.  Moreover,  $\Z_2^n$-gradings appear in the context of parasupersymmetry, see for example  Yang \& Jing \cite{Yang:2001}  and Yang, Jing \& Ping  \cite{Yang:2001b}; can be found behind the symmetries of the L\'{e}vy--Leblond equation, see Aizawa \& Segar \cite{Aizawa:2017}; and Tolstoy's alternative super-Poincar\'{e} symmetries \cite{Tolstoy:2014}. \par 
Technically we enter the world of  $\Z_{2}^{n}$-manifolds (or coloured supermanifolds) which informally are `manifolds' for which the structure sheaf has a  $\Z_{2}^{n}$-grading. The $n=1$ case is just the standard theory of supermanifolds. One can view supermanifolds as a mild form of noncommutative geometry, and similarly one can view $\Z_{2}^{n}$-manifolds as a mild form of `noncommutative supergeometry'. However, as almost all of the differential geometry of supermanifolds and $\Z_{2}^{n}$-manifolds can be treated an interpreted using the methods from classical differential geometry and algebraic geometry. Thus, one should view $\Z_{2}^{n}$-manifolds as a starting place for more general noncommutative supergeometries where more abstract algebraic methods are needed (see for example de~Goursac \cite{Goursac:2015}, Grosse \& Reiter \cite{Grosse:2000} and Schwarz \cite{Schwarz:2003}). \par
 In this paper, we will build, via a $\Z_{2}^{n}$-graded generalisation of the super-Poincar\'{e} algebra,  a $\Z_{2}^{n}$-graded version of super-Minkowski space-time. The methods employed are minor modifications of the standard methods of coset spaces as applied to standard supersymmetry. The aim of this paper is to point out that a rather direct generalisation of extended supersymmetry to the setting of $\Z_2^n$-graded geometry exists. Our methodology is to start from a  $\Z_2^n$-graded generalisation of a  Haag-{\L}opusza\'{n}ski-Sohnius type algebra \cite{Haag:1975} and formally integrate it to construct a $\Z_2^n$-manifold (see for example \cite[Section IV]{Wess:1992} for the standard case). The constructions formally look very similar to the standard case of $\mathcal{N}$-extended supersymmetry. In part, this is why the starting $\Z_2^n$-Lie algebra was chosen as is it. 

\noindent \textbf{Arrangement.} In section \ref{sec:Background} we review the locally ringed space approach to $\Z_{2}^{n}$-manifolds and establish our conventions with spinors.  In section \ref{sec:ZnSupersymmetry} we start from a $\Z_{2}^{n}$-extended Poincar\'{e} algebra, build the corresponding $\Z_{2}^{n}$-Minkowski space-time, and consider some basic consequences of these constructions. Due to the fact that for the $n=2$ case some of the expressions greatly simplify, we present further details of $\Z_{2}^{2}$-Minkowski space-time including its canonical $\Z_2^2$-SUSY structure.   We end this paper with some concluding remarks in section \ref{sec:conclusion}.

\section{Preliminaries}\label{sec:Background}
\subsection{$\Z_{2}^{n}$-manifolds and their basic geometry}
We bring the readers attention to \cite{Covolo:2016,Covolo:2016a, Covolo:2016b, Covolo:2016c} where  details of the locally ringed space approach to $\Z_{2}^{n}$-manifolds can be found. Prior to these works is \cite{Molotkov:2010}, where the functor of points approach was used to define $\Z_{2}^{n}$-manifolds or coloured supermanifolds. We also draw the readers attention to Marcinek \cite{Marcinek:1991}.   We will draw upon these works heavily and not present proofs of any formal statements. Moreover, we will largely follow standard notation from the theory of supermanifolds. We restrict our attention to real structures and do not consider the complex analogues  at all.\par 
\begin{definition}[\cite{Covolo:2016}]
A \emph{locally} $\Z_{2}^{n}$-\emph{ringed space}, $n \in \mathbb{N} \setminus \{0\}$, is a pair $X := (|X|, \mathcal{O}_{X} )$ where $|X|$ is a second-countable Hausdorff topological space and a $\mathcal{O}_{X}$  is a sheaf  of $\Z_{2}^{n}$-graded, $\Z_{2}^{n}$-commutative associative unital $\mathbb{R}$-algebras, such that the stalks $\mathcal{O}_{X,p}$, $p \in  |X|$ are local rings.
\end{definition}
In this context,  $\Z_{2}^{n}$-commutative means that any two sections $a$, $b \in \mathcal{O}_X(|U|)$, $|U| \subset |X|$ open, of homogeneous degree $\deg(a)$ and $\deg(b) \in \Z_{2}^{n}$, respectively, commute with the sign rule
$$ab = (-1)^{\langle \deg(a), \deg(b)\rangle} \: ba,$$
where $\langle \: , \:\rangle$ is the standard scalar product on $\Z_{2}^{n}$.\par 
One should, of course, be reminded of the definition of a superspace, and indeed standard superspaces are examples of locally $\Z_{2}^{n}$-ringed space for $n=1$. To pass from a general superspace to a supermanifold one needs the standard local model of $\mathbb{R}^{p|q}$. The definition of a $\Z_{2}^{n}$-manifold is similar to that of a supermanifold, but with some important subtle differences.\par 
First, we need to fix a convention on how we fix the order of elements in $\Z_{2}^{n}$, we do this \emph{lexicographically} by filling in zeros from the left and ones from the right. We will refer to this ordering as the \emph{standard ordering}. For example, with this choice of ordering
\begin{align*}
 \Z_{2}^{2}  && & = &&  \big \{ (0,0),  \: (0,1), \: (1,0), \: (1,1) \big\}\,,\\
 \Z_2^3  && &=&&   \big \{  (0, 0, 0), \: (0, 0, 1), \:(0, 1, 0), \:(1, 0, 0), \: (0, 1, 1), \:(1, 0, 1), \:  (1,1, 0),\: (1, 1, 1)  \big \}\,, \\
 \Z_2^4  && &=&&    \big \{  
 (0, 0, 0, 0), \: (0, 0, 0, 1),\:  (0, 0, 1, 0),\: (0, 1, 0, 0),\: (1, 0, 0, 
  0), \\
 &&  & &&   (0, 0, 1, 1),\:  (0, 1, 0, 1), \:(0, 1, 1, 0),\:  (1, 0, 0, 1),\: (1, 0, 
  1, 0), \:  (1, 1, 0, 0), \\ 
&&   && &    (0, 1, 1, 1), \:  (1, 0, 1, 1),\:  (1, 1, 0, 1)\:, (1, 1, 1, 0), \:  (1, 1, 1, 1) \big  \}\,. 
\end{align*}
Note that slightly different conventions with the ordering have been employed in the literature. A tuple $\mathbf{q} = (q_{1}, q_{2}, \cdots , q_{N})$, where  $N = 2^{n}-1$ provides all the information about the non-zero degree coordinates, which we collectively write as $\zx$. For example, specifying $\mathbf{q} = \{2,3,6 \}$ means that we have 2 coordinates of degree $(0,1)$, 3 of degree $(0,1)$, and 6 of degree $(1,1)$. 
\begin{definition}[\cite{Covolo:2016}]
A (smooth) $\Z_{2}^{n}$-\emph{manifold} of dimension $p |\mathbf{q}$ is a locally $\Z_{2}^{n}$-ringed space $ M := \left(|M|, \mathcal{O}_{M} \right)$, which is locally isomorphic to the $\Z_{2}^{n}$-ringed space $\mathbb{R}^{p |\mathbf{q}} := \left( \mathbb{R}^{p}, C^{\infty}_{\R^p}[[\zx]] \right)$. Here   $C^{\infty}_{\R^p}$ is the structure sheaf on the Euclidean space $\R^p$.  Local sections of $\mathbb{R}^{p |\mathbf{q}}$ are formal power series in the $\Z_{2}^{n}$-graded variables $\zx$ with  smooth coefficients, i.e.,
$$   C^{\infty}(\mathcal{U}^p)[[\zx]] =  \left \{ \sum_{\alpha \in \mathbb{N}^{N}}^{\infty}  \zx^{\alpha}f_{\alpha} ~ | \: f_{\alpha} \in C^{\infty}(\mathcal{U}^p)\right \},$$
for any  $\mathcal{U}^p \subset \R^p$ open.  \emph{Morphisms} between $\Z_{2}^{n}$-manifolds are  morphisms of $\Z_{2}^{n}$-ringed spaces, that is,  pairs $(\phi, \phi^{*}) : (|M|, \mathcal{O}_{M}) \rightarrow  (|N|, \mathcal{O}_{N})$ consisting of a continuous map  $\phi: |M| \rightarrow |N|$ and sheaf morphism $\phi^{*}: \mathcal{O}_{N}(|V|) \rightarrow \mathcal{O}_{M}(\phi^{-1}(|V|))$, where $|V| \subset |N|$  is open.
\end{definition}
\begin{example}[The local model]
The locally $\Z_{2}^{n}$-ringed space $\mathcal{U}^{p|\mathbf{q}} :=  \big(\mathcal{U}^p , C^\infty_{\mathcal{U}^p}[[\zx]] \big)$, where $\mathcal{U}^p \subseteq \R^p$, is naturally a $\Z_2^n$-manifold, we refer to such $\Z_2^n$-manifolds as \emph{$\Z_2^n$-superdomains} of dimension $p|\mathbf{q}$.  We can employ (natural) coordinates $(x^a, \zx^i)$ on any $\Z_2^n$-superdomain, where $x^a$ form a coordinate system on $\mathcal{U}^p$ and the $\zx^i$ are formal coordinates. 
\end{example}
Canonically associated with any $\Z_{2}^{n}$-graded algebra $\mathcal{A}$ is the homogeneous ideal $J$ of $\mathcal{A}$ generated by
all homogeneous elements of $\mathcal{A}$ having non-zero degree. If $f : \mathcal{A} \rightarrow \mathcal{A}^{\prime}$ is a morphism of $\Z_{2}^{n}$ -graded algebras, then $f (J_{\mathcal{A}} ) \subseteq J_{\mathcal{A}^{\prime}}$. The $J$-adic topology plays a fundamental r\^ole in the theory of $\Z_{2}^{n}$-manifolds. In particular, these notions can be `sheafified'. That is, for any $\Z_{2}^{n}$-manifold $M$, there exists an ideal sheaf $\mathcal{J}$, defined by
$\mathcal{J} (|U| ) = \left \langle f\in \mathcal{O}_{M}(|U|)\: | \: f \: \textnormal{is of non-zero degree}   \right\rangle$. The $\mathcal{J}$-adic topology on $\mathcal{O}_{M}$ can then be defined in an obvious way (see \cite{Covolo:2016} for details).\par  
Many of the standard results from the theory of supermanifolds pass over to $\Z_{2}^{n}$-manifolds. For example, the topological space $|M|$ comes with the structure of a smooth manifold of dimension $p$, hence our suggestive notation. There exists a canonical projection $\epsilon : \mathcal{O}_{M}(|M|) \rightarrow C^{\infty}(|M|)$, and it can be shown that $\mathcal{J}= \ker \epsilon$.\par 
The immediate problem with $\Z_{2}^{n}$-manifolds is that $\mathcal{J}$ is \emph{not} nilpotent -- for standard supermanifolds the ideal sheaf  is nilpotent and this is a fundamental property that makes the theory of supermanifolds so well-behaved. However, this loss of nilpotency is replaced by the Hausdorff completeness of the $\mathcal{J}$-adic topology.
\begin{proposition}[\cite{Covolo:2016}] Let $M$ be a $\Z_{2}^{n}$-manifold. Then $\mathcal{O}_{M}$ is $\mathcal{J}$-adically Hausdorff complete as a sheaf of  $\Z_{2}^{n}$-commutative rings, i.e., the morphism
$$\mathcal{O}_{M} \rightarrow \lim_{\leftarrow k \in \mathbb{N}} \mathcal{O}\setminus \mathcal{J}^{k}$$
naturally induced by the filtration of $\mathcal{O}_{M}$ by the powers of $\mathcal{J}$ is an isomorphism.
\end{proposition}
 The presence of formal power series in the coordinate rings of $\Z_{2}^{n}$-manifolds forces one to rely on the  Hausdorff-completeness of the $\mathcal{J}$-adic topology. This completeness replaces the standard fact that on supermanifolds functions of Grassmann odd variables are always polynomials -- a result that is often used in extending results from manifolds to supermanifolds. \par 
What makes $\Z_{2}^{n}$-manifolds a very workable form of noncommutative geometry is the fact that we have well-defined local models and that the structure of morphisms can (locally) be described at the level of coordinates. That is, we have the \emph{chart theorem} (\cite[Theorem 7.10]{Covolo:2016}) that basically says that morphisms between the local coordinates can be extended uniquely to morphisms of locally $\Z_{2}^{n}$-ringed spaces. \par 
Let $M$ be a $\Z_{2}^{n}$-manifold, then an \emph{open $\Z_{2}^{n}$-submanifold} of $M$ is a  $\Z_{2}^{n}$-manifold defined as 
$$U := \big( |U|, \cO_{M,|U|} \big),$$
where $|U| \subseteq |M|$ is an open subset. From the definition of a $\Z_2^n$-manifold, we know that for `small enough' $|U|$ we have an isomorphism
$$ (\varphi, \varphi^*): U  \stackrel{\sim}{\rightarrow} \mathcal{U}^{p|\mathbf{q}}\,.$$
This local identification allows us to construct a local coordinate system.  Dropping explicit reference to the local isomorphism, and so via minor abuse of notation, we write the local coordinates  as $x^{A} = (x^{a}, \zx^{i})$.  The commutation rules for these coordinates is determined by the  scalar product on $\Z_{2}^{n}$ inherited by the standard (Euclidean) scalar product on $\R^{n}$ ,i.e.,
$$x^{A}x^{B} = (-1)^{\langle \deg(A), \deg(B)\rangle}\:x^{B}x^{A}\,.$$

Changes of coordinates, i.e., different choices of the local isomorphisms,  can be written (using standard abuses of notation) as $x^{A'} = x^{A'}(x)$, where we understand the changes of coordinates to respect the $\Z_{2}^{n}$-grading. The subtlety here is that these changes of coordinates need not be polynomial in coordinates of non-zero degree, generically we have a formal power series. We will refer to global sections of the structure sheaf of a $\Z_{2}^{n}$-manifold as \emph{functions} and employ the standard notation  $ C^{\infty}(M):=\cO_M(|M|)$. 
\begin{remark}
It is clear that, much like standard supermanifolds,  a  $\Z_{2}^{n}$-manifold is not `simply' a collection of topological points. The only topological points of $M$ are the points of the reduced manifold $|M|$. To regain some classical intuition one can employ Grothendieck's functor of points. However, we will not need to do so in this work. Or rather, we avoid using the functor of points and work formally in the rare instances we need to - principally in defining group structures on $\Z_{2}^{n}$-manifolds.
\end{remark}
\begin{example}[$\Z_2^n $-graded  Cartesian spaces] Directly from the definition, it is clear that  $\mathbb{R}^{p |\mathbf{q}} := \left( \mathbb{R}^{p}, C^{\infty}(\mathbb{R}^{p})[[\zx]] \right)$ is a $\Z_2^n $-manifold. Global coordinates $(x^a, \zx^i)$ can be employed, where the coordinate map is just the identity. In this paper, we will only meet $\Z_2^n$-manifolds that are globally isomorphic to $\R^{p|\mathbf{q}}$ for the appropriate $p$ and $\mathbf{q}$.
\end{example} 
\begin{example}[Manifolds and supermanifolds]
Rather trivially, smooth manifolds are $\Z_{2}^{0}$-manifolds.  Similarly, supermanifolds are $\Z_{2}^{1}$-manifolds. Note that for supermanifolds, Grassmann odd coordinates are nilpotent, and thus functions are polynomial in odd variables.
\end{example}
\begin{example}[Double vector bundles] A double vector bundle $D$ is manifold that has a pair of commuting actions of the multiplicative  monoid of reals: $\rmh_{i} : \mathbb{R} \times D \rightarrow D$ $(i = 1,2)$. This amounts to being able to find homogeneous (with respect to the natural bi-weight) local coordinates (we suppress indices)
$$(\underbrace{x}_{(0,0)},\: \underbrace{y}_{(0,1)},\: \underbrace{z}_{(1,0)},\: \underbrace{w}_{(1,1)}),$$
see \cite{Grabowski:2009,Voronov:2001}. Admissible changes of homogeneous local coordinates are of the form
\begin{align*}
& x' = x'(x), & y' = y S(x), && z' = zT(x) , && w' = wU(x) + zy V(x).
\end{align*}
Given  any double vector bundle, one can canonically associate, in a functorial way, a $\Z_{2}^{2}$-manifold by declaring the commutation rules for the local coordinates to be determined by the bi-weight. The changes of coordinates remain as given above.  For more details consult \cite[Section 6]{Covolo:2016a}, where the resulting  $\Z_{2}^{2}$-manifold is denoted $\Pi D$.
\end{example}
The notion of a vector field on a $\Z_{2}^{n}$-manifold is naturally given as a $\Z_{2}^{n}$-derivation acting on sections of the structure sheaf, i.e., acting on (global) functions. A (homogeneous) vector field is understood as  a linear map $X: C^{\infty}(M) \rightarrow C^{\infty}(M)$, that satisfies the $\Z_2^n$-graded Leibniz rule
$$X(fg) = X(f) g + (-1)^{\langle \deg(X), \deg(f) \rangle}\: f X(g),$$
for any $f$ and $g \in C^{\infty}(M)$. It is a simple exercise to see that we have a (left) module structure over $C^{\infty}(M)$. We denote the module of vector fields as $\Vect(M)$. Any vector field can be `localised' (see \cite[Lemma 2.2]{Covolo:2016b}) in the sense that given $|U| \subset |M|$ there always exists a unique derivation
$$X|_{|U|} : \cO_M(|U|) \rightarrow \cO_M(|U|),$$
such that $X(f)|_{|U|} = X|_{|U|}(f_{|U|})$. Because of this local property, it is clear that one has a sheaf of $\cO_M$-modules formed by the local derivations -- this defines the \emph{tangent sheaf} of a $\Z_{2}^{n}$-manifold (see \cite[Definition 5.]{Covolo:2016b}). Moreover, this sheaf is locally free and so admits a local basis. The upshot of these considerations is that we can always locally write a vector field as one would on a manifold using a choice of local coordinates, i.e.,
$$X = X^{A}(x)\frac{\partial}{\partial x^{A}},$$
where the partial derivatives are defined as standard for the coordinates of degree zero and are defined algebraically for the coordinates of degree non-zero. In the above expression, we drop the explicit reference to the required restriction as is common in standard differential geometry.  As for the case of standard supermanifolds, the order of taking partial derivatives matters, but only up to signs,
$$\frac{\partial}{\partial x^{A}} \frac{\partial}{\partial x^{B}} =  (-1)^{\langle \deg(A), \deg(B)  \rangle } \: \frac{\partial}{\partial x^{B}} \frac{\partial}{\partial x^{A}}.$$
Under the commutator 
$$[X,Y] :=  X \circ Y - (-1)^{\langle \deg(X), \deg(Y)\rangle } \:  Y\circ X,$$
$\Vect(M)$ becomes a $\Z_{2}^{n}$-Lie algebra (see \cite{Covolo:2012,Rittenberg:1978,Scheunert:1979}). The  grading and symmetry  of the Lie bracket are clear, and one can check that the Jacobi identity  (written here in Loday--Leibniz form)
$$[X,[Y,Z]] = [[X,Y],Z] + (-1)^{\langle \deg(X), \deg(Y) \rangle } \: [Y, [X,Z]],$$
holds. \par 
Any geometric object -- say a function, vector field or tensor field -- on a $\Z_{2}^{n}$-manifold carries a natural $\Z_{2}^{n}$-grading. We will say that a homogeneous geometric object is \emph{even} if the total degree is even, and \emph{odd} if the total degree is odd.  Note that this does \emph{not} determine the sign factors in general expressions.

\begin{remark}
 It is known that the structure sheaf of a $\Z_2^n$-manifold is a nuclear Fr\'{e}chet sheaf of $\Z_2^n$-graded $\Z_2^n$-commutative associative unital algebras. Moreover, like  real manifolds and supermanifolds, the structure sheaf can be fully recovered from the algebra of global functions. For example, it can be shown that two $\Z_2^n$-manifolds are diffeomorphic if and only if their algebras of global functions are isomorphic. Let us also mention that the category of $\Z_2^n$-manifolds admits finite products - an essential fact in any rigours treatment of $\Z_2^n$-Lie groups. None of this will feature in the later sections of this paper.  For details, the reader should consult Bruce \&  Poncin \cite{Bruce:2018a, Bruce:2018b}.
\end{remark}
\subsection{A toy $\Z_2^2$-superspace}
Before giving a more careful description of $\Z_2^n$-Minkowski space-time we proceed with a simpler toy superspace in which we have an underlying one-dimensional space-time. This will allow us to sketch the ideas without the clutter of many indices and in particular spinor indices.  Let us consider the $\Z_2^2$-graded Lie algebra with generators $P, Q_1, Q_2$ and $Z$ of $\Z_2^2$-degrees $(0,0)$, $(0,1)$, $(1,0)$  and $(1,1)$, respectively, given by
\begin{align*}
& [Q_1,Q_1] = [Q_2, Q_2] = \frac{1}{2}P, && [Q_2,Q_1] = \frac{1}{2}Z\,,
\end{align*}
where all other Lie brackets vanish.  Up to conventions, this is essentially two copies of supersymmetry in one-dimension and a central extension. However, one must take care as we have $\Z_2^2$-graded Lie brackets, the first two brackets are `anticommutators' while the third is a `commutator' in `non-graded language'. We want to realise these generators as vector fields on some $\Z_2^2$-manifold, or more correctly we are looking the  associated  $\Z_2^2$-Lie group and the left-invariant vector fields.  In this case, we can more-or-less see by eye, i.e., without resorting to coset methods,  what the $\Z_2^2$-manifold and left-invariant vector fields should be. With this in mind, consider the $\Z_2^2$-manifold, which we denote as $\mathcal{M}_1^{[2]} \simeq \R^{1|1,1,1}$, that comes with global coordinates
$$\big ( \underbrace{t}_{(0,0)}, \:  \underbrace{\theta^1}_{(0,1)}, \: \underbrace{\theta^2}_{(1,0)}, \: \underbrace{z}_{(1,1)}  \big)\,,$$
where again we have used the standard ordering.  In comparison with the case of standard superspace note the unusual commutation relations
\begin{align*}
& \theta^1 \theta^2 =  {+} \theta^2 \theta^1, && z \theta^1 = {-} \theta^1 z, & z \theta^2 = {-} \theta^2 z\,.
\end{align*}
The reader can easily check that the vector fields
\begin{align*}
& P = \frac{\partial}{\partial t},
& Q_1 = \frac{\partial}{\partial \theta^1} + \frac{1}{4} \theta^1 \frac{\partial}{\partial t} + \frac{1}{4}\theta^2 \frac{\partial}{\partial z},
&& Q_2 = \frac{\partial}{\partial \theta^2} + \frac{1}{4} \theta^2 \frac{\partial}{\partial t}{ -} \frac{1}{4}\theta^1 \frac{\partial}{\partial z},
&& Z = \frac{\partial}{\partial z},
\end{align*}
satisfy the given $\Z_2^2$-Lie algebra under standard $\Z_2^2$-commutators. From these vector fields, we can then `read off' the ($d=1$) \emph{$\Z_2^2$-supersymmetry transformations}
\begin{align*}
& t \mapsto t + \frac{1}{4} \big( \epsilon^1 \theta^1 + \epsilon^2 \theta^2 \big),
&& \theta^1 \mapsto \theta^1 + \epsilon^1,
&& \theta^2 \mapsto \theta^2 + \epsilon^2,
&&  z \mapsto z + \frac{1}{4}\big( \epsilon^1 \theta^2 \:  {-} \:\epsilon^2 \theta^1 \big),
\end{align*}
where $\epsilon^1$ and $\epsilon^2$ are $\Z_2^2$-graded parameters of degree $(0,1)$ and $(1,0)$, respectively. It is a direct calculation to derive the covariant derivatives, which can easily be shown to be given by:
\begin{align*}
& D_1 = \frac{\partial}{\partial \theta^1} {-} \frac{1}{4} \theta^1 \frac{\partial}{\partial t} {-} \frac{1}{4} \theta^2 \frac{\partial}{\partial z}\,, && D_2 = \frac{\partial}{\partial \theta^2} {-} \frac{1}{4} \theta^2 \frac{\partial}{\partial t} {+} \frac{1}{4}\theta^1 \frac{\partial}{\partial z}\,.
\end{align*}
As fully expected, we have $[D_1, D_1] = [D_2, D_2] = - \half P$ and $[D_2 , D_1] = - \half Z$,  and all the other possible  $\Z_2^2$-commutators vanish. \par 
In order to construct  more `realistic' $\Z_2^n$-superspaces we need to take the underlying space to be Minkowski space-time and promote the odd coordinates to be spinors. We will discuss the specific example of  $d=4$ $\Z_2^2$-Minkowski space-time in subsection \ref{subsec:Z2Minkowski}.\par 
To understand the `field content' encoded here let us first remove the central charge, i.e., we set $z =0$, as this will simplify things - otherwise we will have to deal with a formal power series in $z$. A (scalar) \emph{superfield} of degree $(0,0)$ is then of the form 
$$\Phi(t, \theta)) = q(t) + \theta^1 \chi_1(t) + \theta^2 \chi_2(t) + 4 \theta^2\theta^1 b(t),$$
where we have included a numerical factor for convenience. Superfields require the use of internal Homs in order to define them rigorously, i.e., we  need to consider some `external parametrisations' so that the degrees can be properly attributed. However, we will simply proceed formally as is common in physics. Specifically, the degrees of the components are
\begin{align*}
& \deg(q) = (0,0), && \deg(\chi_1) = (0,1), && \deg(\chi_2) = (1,0), && \deg(b) = (1,1).
\end{align*}
Clearly, the induced commutation rules between these components are different to what is encountered in standard supersymmetric mechanics, see for example Bellucci \& Krivonos \cite{Bellucci:2006}.  In particular, we have odd degrees of freedom that commute with each other  and anticommute with one of the even degrees of freedom. At this stage the reader should be reminded of parastatistics in the sense of Green \cite{Green:1953}. We define the component form of the $\Z_2^2$-supersymmetry transformations via $\delta \Phi = (\epsilon^1 Q_1 + \epsilon^2 Q_2 )\Phi$. It is a straightforward exercise to deduce the following:
\begin{align*}
& \delta q(t) = \epsilon^1 \chi_1(t) + \epsilon^2 \chi_2(t), && \delta \chi_1(t) = - \frac{1}{4} \big( \epsilon^1 \dot q(t) - \epsilon^2 b(t) \big),\\ & \delta \chi_2(t) = - \frac{1}{4} \big( \epsilon^2 \dot q(t) - \epsilon^1 b(t) \big), && \delta b(t) = - \big( \epsilon^1 \dot{\chi}_2(t) + \epsilon^2 \dot{\chi}_1(t) \big) ,
\end{align*}
where `dot' means the derivative with respect to $t$. These transformations should be compared to the (off-shell) $d= 1$ $\mathcal{N}=2$ supersymmetry transformations, see for example \cite[Section 2.2]{Bellucci:2006}. The reader should note that we work in a manifestly real setting and so no factors of  the imaginary unit  appear in any of our mathematical expressions.\par 
Via dimensional reduction, for $d=0$ the $\Z_2^2$-graded Lie algebra we should consider is simply
\begin{align*}
& [Q_1,Q_1] = [Q_2, Q_2] = 0, && [Q_2,Q_1] = \frac{1}{2}Z\,,
\end{align*}
where the $\Z_2^2$-degrees are assigned as previously. Thus, $\mathcal{M}_0^{[2]} \simeq \R^{0|1,1,1}$ comes with vector fields
\begin{align*}
& Q_1 = \frac{\partial}{\partial \theta^1}+ \frac{1}{4}\theta^2 \frac{\partial}{\partial z},
&& Q_2 = \frac{\partial}{\partial \theta^2} { -} \frac{1}{4}\theta^1 \frac{\partial}{\partial z},
&& Z = \frac{\partial}{\partial z},
\end{align*}
that realise the  $d=0$, $\Z_2^2$-supersymmetry algebra. Note that in this case $Q_1$ and $Q_2$ are \emph{homological}, i.e., they square to zero.  We do not obtain a bi-complex in this way as the two homological vector fields to not commute, unless the central charge $Z=0$. 

\subsection{Majorana spinors}
We take the Minkowski metric to be $\eta = \textnormal{diag}(-1,+1, +1 ,+1)$ and  denote  3+1 dimensional Minkowski space-time as $\cM_4 := (\R^{4}, \eta)$. That is, we will use the relativists' conventions.  We wish to work with \emph{real} objects rather than complex ones. As such  will choose the following \emph{real Majorana representation} of the Clifford algebra $\mathcal{C}l(3,1)$
\[ \gamma^{0} = \left( \begin{array}{rrrr}
0 & +1 & 0 & 0 \\
-1 & 0 & 0 & 0  \\
0 & 0 & 0  & -1\\
0 & 0 & +1  & 0\\
\end{array} \right), \hspace{25pt}
 \gamma^{1} = \left( \begin{array}{rrrr}
0 & +1 & 0 & 0 \\
+1 & 0 & 0 & 0  \\
0 & 0 & 0  & +1\\
0 & 0 & +1  & 0\\
\end{array} \right),\]
\\
\[ \gamma^{2} = \left( \begin{array}{rrrr}
+1 & 0 & 0 & 0 \\
0 & -1 & 0 & 0  \\
0 & 0 & +1  & 0\\
0 & 0 & 0  & -1\\
\end{array} \right), \hspace{25pt}
 \gamma^{3} = \left( \begin{array}{rrrr}
0 & 0 & 0 & +1 \\
0 & 0 & -1 & 0  \\
0 & -1 & 0  & 0\\
+1 & 0 & 0  & 0\\
\end{array} \right).\]
These gamma matrices -- by definition --  satisfy the \emph{Clifford-Dirac relation}
$$\{\gamma^{\mu}, \gamma^{\nu}  \}:= \gamma^{\mu} \gamma^{\nu} + \gamma^{\nu} \gamma^{\mu} = 2 \eta^{\mu \nu}\Id .$$\par 
These $4\times 4$ real matrices act on $\R^{4}$ in the usual way. With this in mind, we adopt the convention of spin indices (and for matrices more generally) of $(\gamma^{\mu})_{\alpha}^{\:\:\: \beta}$. We then define \emph{Majorana spinors} as the ``things" that these gamma matrices act on.  We thus write $\mathbb{M} \simeq \R^{4}$ for the space of all Majorana spinors. We will conventionally choose the components if a Majorana spinor to have \emph{lower spin indices}, and so we write $u_{\alpha}$ for a Majorana spinor. We know that $\textnormal{Spin}(3,1)$ acts on Majorana spinors in a linear way, i.e.,
$$u_{\alpha} \mapsto u'_{\alpha} = S_{\alpha}^{\:\:\: \beta}u_{\beta}.$$
\noindent Given an infinitesimal Lorentz transformation
$$\Lambda_{\mu}^{\:\:\: \nu} = \delta_{\mu}^{\:\:\: \nu} + \omega_{\mu}^{\:\:\:\nu},$$
where $\omega_{\mu \nu} = - \omega_{\nu \mu}$, we know that a Majorana spinor transforms as
$$u'_{\alpha} = u_{\alpha} + \frac{1}{4} \omega_{\mu \nu}(\gamma^{\mu \nu})_{\alpha}^{\:\:\: \beta}u_{\beta},$$
where $\gamma^{\mu \nu} :=  \gamma^{[\mu} \gamma^{\nu ]} =  \half( \gamma^{\mu} \gamma^{\nu }  -  \gamma^{\nu} \gamma^{\mu })$. In particular, $\Sigma^{\mu \nu} = \half\gamma^{\mu \nu}$ give a representation of the \emph{Lorentz algebra}
$$[\Sigma^{\mu \nu}, \Sigma^{\rho \sigma}] = \eta^{\nu \rho}\Sigma^{\mu \sigma} -  \eta^{\mu \rho}\Sigma^{\nu \sigma} + \eta^{\mu \sigma}\Sigma^{\nu \rho} - \eta^{\nu \sigma}\Sigma^{\mu \rho}.$$
By taking the exponential we obtain a finite Lorentz transformation (which by construction is in the proper Lorentz group)
$$u'_{\alpha} = \exp( \frac{1}{4} \omega_{\mu \nu}\gamma^{\mu \nu} )_{\alpha}^{\:\:\: \beta} u_{\beta}.$$
As defined here, the Majorana spinor as commuting objects: they are  described by the coordinates on $\mathbb{M}$ considered as a linear manifold. In order to have anticommuting Majorana spinors -- which are more common in physics -- we need to employ the parity reversion functor. Thus, we consider anticommuting Majorana spinors to be described by the coordinates on $\Pi \mathbb{M}$. The manifold $\mathbb{M}$ (and the supermanifold $\Pi \mathbb{M}$) comes equipped with  the \emph{charge conjugation tensor} --  this operator exchanges particles and antiparticles. The defining property is
$$C \gamma^{\mu}C^{-1} = - (\gamma^{\mu})^{\textnormal{t}}.$$
Or written more explicitly
$$  C^{\alpha \gamma}(\gamma^{\mu})_{\gamma}^{\:\: \delta} C_{\delta \beta} = {-}(\gamma^{\mu})^{\alpha}_{\:\: \beta},$$
Thus, considered as a matrix,  in our chosen representation, $ C =  -\gamma^{0}$. To set some useful notation, we define $(C\gamma^{\mu})^{\alpha \beta} :=  C^{\alpha \delta }(\gamma^{\mu})_{\delta}^{\:\: \beta}$ and $(C\gamma^{\mu\nu})^{\alpha \beta} :=  C^{\alpha \delta }(\gamma^{\mu \nu})_{\delta}^{\:\: \beta}$, both of which a  direct computation shows are symmetric in $\alpha$ and $\beta$.  
\begin{remark}
In practice, we will not need the explicit representation  of $\mathcal{C}l(3,1)$  that we have chosen. The key point is that we have a completely real representation. Furthermore, real representations are only possible in the `mostly pluses' convention for the Minkowski metric.
\end{remark}
\begin{remark}\label{rem:DiracWeyl}
\emph{Dirac spinors} are understood as the complexification of the Majorana spinors, that is we define the space of Dirac spinors as  $\mathbb{D} :=  \mathbb{M} \otimes_{\R} \C \simeq \C^4$. \emph{Weyl spinors} are then defined in terms of a decomposition of the Dirac spinors into a direct sum $\mathbb{D} = \mathbb{W}_{+}\oplus \mathbb{W}_{-}$ defined by chirality.  More explicitly, we first define $\gamma_{5}:= \gamma_{0}\gamma_{1}\gamma_{2}\gamma_{3}$ (where we have used the Minkowski metric to lower the indices). The decomposition is then defined in terms of the eigenvalues of $\gamma_{5}$. That is,  $u \in \mathbb{W}_{\pm}$ if  $\gamma_{5} u = \pm \rmi \: u$. Via this decomposition, we see that $\mathbb{W}_{\pm} \simeq \C^2$. We will only consider Majorana spinors in this paper.
\end{remark}
\subsection{$\Z_{2}^{n}$-graded Majorana spinors}
We can construct $\Z_{2}^{n}$-graded Majorana spinors quite directly. From the Batchelor--Gaw\c{e}dzki theorem for $\Z_{2}^{n}$-manifolds \cite[Theorem 3.2]{Covolo:2016a} we know that any (real) $\Z_{2}^{n}$-manifold is noncanonically isomorphic to a  $\Z_{2}^{n}\setminus \{\underline{0} \}$-graded vector bundle. In particular, any $\Z_{2}^{n}\setminus \{\underline{0} \}$-graded manifold is noncanonically isomorphic to the direct sum of the appropriate number of vector spaces -- where we assign the appropriate $\Z_{2}^{n}\setminus \{\underline{0} \}$-grading to the linear coordinates on each vector space. Each vector space is then considered as a linear  $\Z_{2}^{n}$-manifold.
\begin{example}
Any $\Z_{2}^{2}\setminus \{\underline{0} \}$-graded manifold is noncanonically isomorphic to a direct sum of vector spaces $E = E_{(0,1)} \oplus E_{(1,0)} \oplus E_{(1,1)}$, where the labels indicate the assignment of the degree to the coordinates on each vector space. Note that as coordinate changes are linear the assignment of the degree is well defined.
\end{example}
Then to define the  space of $\Z_{2}^{n}$-graded Majorana spinors we simply take the appropriate number of copies of $\mathbb{M}$, assign the degrees  to the linear coordinates, and then  take the direct sum. To set notation, we denote the resulting $\Z_{2}^{n}$-manifold as $\mathbb{M}^{[n]}$.
\begin{example}
The $n=1$ case is the classical case of anticommuting spinors, $\mathbb{M}^{[1]} = \Pi \mathbb{M}$.
\end{example}
\begin{example}
The $n=2$ case is as follows:  $\mathbb{M}^{[2]} = \mathbb{M}_{(0,1)}\oplus \mathbb{M}_{(1,0)} \oplus \mathbb{M}_{(1,1)}$. In terms of coordinates, we have three `species' of Majorana spinors $(\psi_{\alpha}, \theta_{\beta}, u_{\gamma})$ of degree $(0,1)$, $(1,0)$ and $(1,1)$ respectively. Note that although $\psi$ and $\theta$ are odd, they \emph{commute} and not anticommmute. Similarly, $u$ are even yet \emph{anticommute} with both  $\psi$ and $\theta$.
\end{example}
In the next section, we will be interested in \emph{odd Majorana spinors}. Note that we have the natural decomposition
$$\mathbb{M}^{[n]} = \mathbb{M}^{[n]}_{0} \oplus \mathbb{M}^{[n]}_{1}, $$
into even and odd spinors defined by the total degree. We will exclusively be interested in $\mathbb{M}^{[n]}_{1}$ in the rest of this paper. It is a simple counting exercise to show that $\mathbb{M}^{[n]}_{1}$ has $4\times 2^{n-1}$ coordinates.


\section{$\Z_{2}^{n}$-extended supersymmetry}\label{sec:ZnSupersymmetry}
\subsection{A $\Z_{2}^{n}$-extended Poincar\'{e} algebra}
We extend the standard Poincar\'{e}  algebra by adjoining $\Z_{2}^{n}$-graded odd Majorana spinor generators $Q_{I}^{\alpha}$ and $\Z_{2}^{n}$-graded even central charges $Z_{IJ}$.   Experimentally we know that the underlying real manifold  of the resulting $\Z_2^n$-manifold must, locally at least, be Minkowski space-time. Thus, the resulting algebra must contain the Poincar\'{e}  algebra as the degree zero component.  The $\Z_{2}^{n}$-graded Lie algebra we consider is a generalisation of the  $N = 2^{n-1}$ Haag-{\L}opusza\'{n}ski-Sohnius type algebra \cite{Haag:1975}, i.e., we allow a central extension:
\begin{align*}
& [P_{\mu} , J^{\lambda \sigma}] = (\delta_{\mu}^{\:\: \lambda} \eta^{\sigma \rho} -\delta_{\mu}^{\:\: \sigma} \eta^{\lambda \rho} )P_{\rho},\\
&[J^{\mu \nu} , J^{\rho  \sigma}] = \eta^{\nu \rho}J^{\mu \sigma} - \eta^{\mu \rho}J^{\nu \sigma}  - \eta^{\nu \sigma}J^{\mu \rho} + \eta^{\mu \sigma}J^{\nu \rho},\\
& [J^{\mu \nu}, Q_{I}^{\alpha}]  = - \frac{1}{4}Q_{I}^{\beta}(\gamma^{\mu \nu})_{\beta}^{\:\: \alpha},\\
& [Q_{I}^{\alpha}, Q_{J}^{\beta}] = \frac{1}{2} \delta_{IJ} (C\gamma^{\mu})^{\alpha \beta}P_{\mu} + C^{\alpha \beta}Z_{IJ} .
\end{align*} 
All other Lie brackets are zero. In particular, $Z$ is central.  Here $P$ and $J$ are the generators of translations and rotations on Minkowski space-time. To be clear, the degree of the generators is as follows:
\begin{align*}
& \deg(P) =  \deg(J) = \underline{0}, && (\textnormal{even}) \\
& \deg(Q_{I}) = \deg(I) \in \Z_{2}^{n}\setminus \{ \underline{0}\}, &&  (\textnormal{odd})\\
& \deg(Z_{IJ}) = \deg(I) + \deg(J) \in \Z_{2}^{n}\setminus \{ \underline{0}\}. && (\textnormal{even})
\end{align*}
From the symmetry of $\Z_2^n$-Lie brackets  and the fact that $C^{\alpha \beta} = {-} C^{\beta \alpha}$, it is clear that $Z_{IJ} = (-1)^{\langle \deg(I) , \deg(J) \rangle} \:Z_{JI}$ for $I \neq J$, and  that $Z_{II} = - Z_{II} =0$. Thus, as $Z_{IJ}$ and $Z_{JI}$ are \emph{not} linearly independent as elements of the  $\Z_{2}^{n}$-graded Lie algebra. \par 
To see that we do indeed have a $\Z_2^n$-Lie algebra we need to establish that the necessary Jacobi identities are satisfied. First, note that all possible Jacobi identities involving just one $Q_I^\alpha$ hold as this situation essentially reduces to the classical $\mathcal{N}=1$ super-Poincar\'{e} algebra. \par
The Jacobi identities involving $Q_I^\alpha$ and  $Q_J^\beta$ require checking. Any nested commutators involving two $Q$ and  one $P$ are identically zero  due to $[Q, Q] = P + Z$ and $[Q,P] = [Q,Z] =0$ (we suppress the indices and coefficients).  The only non-trivial identity to check is
$$[Q_I^\alpha , [J^{\mu \nu}, Q_J^\beta]] \stackrel{?}{=} [[Q_I^\alpha, J^{\mu \nu}] , Q_J^\beta]  + [J^{\mu \nu} , [Q_I^\alpha , Q_J^\beta]].$$  
If $I = J$ then we are back to the standard $N=1$ super-Poincar\'{e} algebra and there is nothing to check. So we assume $I \neq J$. Directly from the algebra  we have
$$[Q_I^\alpha , [J^{\mu \nu}, Q_J^\beta]] = - \frac{1}{4}C^{\alpha \gamma}(\gamma^{\mu \nu})_{\gamma}^{\:\: \beta}Z_{IJ} = - \frac{1}{4}(C\gamma^{\mu \nu})^{\alpha \beta}Z_{IJ}.$$
From the other side 
\begin{align*}
[[Q_I^\alpha, J^{\mu \nu}] , Q_J^\beta]  + [J^{\mu \nu} , [Q_I^\alpha , Q_J^\beta]] &= -[ [J^{\mu \nu}, Q_{I}^\alpha], Q_{J}^\beta] + [J^{\mu \nu }, C^{\alpha \beta} Z_{IJ}]\\
 & = \frac{1}{4}C^{\gamma \beta}(\gamma^{\mu \nu})_{\gamma}^{\:\: \alpha}Z_{IJ}\\
 &= - \frac{1}{4}(C\gamma^{\mu \nu})^{\beta \alpha}Z_{IJ},
 \end{align*}
where we have used $C^{\alpha \beta } = {-}C^{\beta \alpha } $. Then as $(C\gamma^{\mu \nu})^{\beta \alpha} = (C\gamma^{\mu \nu})^{ \alpha \beta} $ we obtain the desired result.\par 
The Jacobi identity involving $Q_I^\alpha$, $Q_J^\beta$ and $Q_K^\gamma$ is trivially satisfied as $[Q, Q] = P + Z$ and $[Q,P] = [Q,Z] =0$ implies that $[Q, [Q,Q]] =0$. Thus, the \emph{$\Z_{2}^{n}$-super translation algebra}, i.e, the part not involving the Lorentz generator $J^{\mu \nu}$, is two-step nilpotent. This is not unexpected given the classical case of the  $\mathcal{N}=1$ super-Poincar\'{e} algebra. It is important to note that, by construction, the Poincar\'{e}  algebra is the degree $\underline{0} = (0,0, \dots, 0)$ part  of the $\Z_{2}^{n}$-extended Poincar\'{e} algebra and not just the even part.  
\begin{remark}
The structures we study are different from the \emph{Poincar\'{e} parasupergroup} of Beckers \& Debergh \cite{Beckers:1993} (also see Nikitin \& Galkin \cite{Nikitin:2000} where central charges were included) and Jarvis \cite{Jarvis:1978}, though we will later comment on the similarities with parasupersymmetry in subsection \ref{subsec:Superfields}.  In particular, the (extended) \emph{Poincar\'{e} parasuperalgebra} is defined in terms of a double commutator for the spinor-valued generators.   The structures we study are $\Z_2^n$-Lie algebras.
\end{remark}

\begin{example}
If $n=1$ then we recover (up to conventions) the standard $\mathcal{N}=1$ super-Poincar\'{e} algebra. Note that we cannot have a non-vanishing central extension in this case.
\end{example}
\begin{example}
Less obvious than the previous example is the case of $n=2$. On the face of it, this looks like the $\mathcal{N}=2$ super-Poincar\'{e} algebra with a central charge. However, the subtlety  is in the assignment of the degree, which are as follows:
\begin{align*}
& \deg(P) = \deg(J)  = (0,0), && \deg(Z_{12}) = (1,1),\\
& \deg(Q_{1}) = (0,1), && \deg(Q_{2}) = (1,0).
\end{align*}
\end{example}
\begin{example}
More complicated again -- though this example highlights the general features -- is the case of $n=3$. Forgetting the obvious assignment of degree to the generators of the Poincar\'{e} algebra, we have
\begin{align*}
& \deg(Q_{1}) =  (0,0,1), && \deg(Z_{12}) = (0,1,1),\\
& \deg(Q_{2}) =  (0,1,0), && \deg(Z_{13}) = (1,0,1),\\
& \deg(Q_{3}) =  (1,0,0), && \deg(Z_{14}) = (1,1,0),\\
& \deg(Q_{4}) =  (1,1,1), && \deg(Z_{23}) = (1,1,0),\\
&                         && \deg(Z_{24}) = (1,0,1),\\
&                         && \deg(Z_{34}) = (0,1,1).\\
\end{align*}
\end{example}
\begin{remark}
We could also include a further modify the extended Poincar\'{e} algebra by including a further term,
$$[Q_{I}^{\alpha}, Q_{J}^{\beta}] = \frac{1}{2} \delta_{IJ} (C\gamma^{\mu})^{\alpha \beta}P_{\mu} + C^{\alpha \beta}Z_{IJ} + (C\gamma_{5})^{\alpha \beta}Y_{IJ},$$
where $\gamma_{5} := \gamma_{0}\gamma_{1}\gamma_{2}\gamma_{3}$. It was shown by  Witten \& Olive \cite{Witten:1978} that in the standard case of $\mathcal{N} \geq 2$ extended supersymmetry that $Z$ and $Y$ have the interpretation as \emph{electric} and \emph{magnetic charges} respectively. Such central charges can arise as the boundary terms in supersymmetric field theories.  One can think of $Z$ in a more general context as colour charges, or even some more exotic force charge. In the case of the $\Z_{2}^{n}$-extended Poincar\'{e} algebra, we have a further problem with interpreting this charge due to the assignment of non-trivial $\Z_{2}^{n}$-degree.   In $(1+1)$ dimensions, a related supersymmetry algebra where only $Y$ appears in the anticommutator of the supersymmetry generators was studied by Duplij, Soroka \&  Soroka \cite{Duplij:2006}.
\end{remark}

\subsection{Direct consequences of the algebra} 
The standard consequences of supersymmetry  upon a quantum theory hold in this higher graded setting. For immediate simplicity let us drop the central term (not that this actually makes much difference). Let us suppose that the  $\Z_{2}^{n}$-super translation algebra (minus central term) can be realised in terms of Hermitian operators acting on some Hilbert space -- the Hilbert space of states of some theory. Then the algebraic structure here implies the following: 
\begin{enumerate}
\item Positivity of energy: a direct computation shows $E := P_{0} = \frac{1}{2}\sum_{\alpha}[Q_{I}^{\alpha}, Q_{I}^{\alpha}]$ (no sum over $I$). Then, passing to the representation as hermitian operators allow us to write $\hat{E} = \sum_{\alpha}(Q_{I}^{\alpha})^{\dag}Q_{I}^{\alpha}$ and thus for any state
$$\langle \psi | \hat{E} | \psi \rangle =\langle \psi | \sum_{\alpha}(Q_{I}^{\alpha})^{\dag}Q_{I}^{\alpha} | \psi \rangle  = ||\sum_{\alpha}Q_{I}^{\alpha} |\psi \rangle ||^{2} \geq 0.$$
\item Irreducible representations of supersymmetry carry the same value of $P^{\mu}P_{\mu} = {-}m^{2}$: this follows from $[P,Q]=0$, which implies that $P^{2}$ is a Casimir.  
\item The spin of each state in a multiplet varies in steps of 1/2: this follows from $[Q,J] \sim Q$.
\end{enumerate}
Thus, there is no difference between the standard super and $\Z_{2}^{n}$-super case with regards to the positivity of energy, Casimirs and the spin of states in each multiplet. However, the notion of BPS states seems more complicated, see subsection \ref{subsec:Z2Minkowski}.

\subsection{$\Z_{2}^{n}$-Minkowski space-time}\label{subsec:ZnMinkowski}
Armed with the theory of $\Z_{2}^{n}$-manifolds, we can more-or-less follow the standard methods from supersymmetry to construct the $\Z_{2}^{n}$-Lie group associated with the $\Z_{2}^{n}$-super translation algebra. That is, we can exponentiate the $\Z_{2}^{n}$-Lie algebra to obtain a $\Z_{2}^{n}$-Lie group. To make rigorous mathematical sense of all these constructions one needs the functor of points -- we will avoid a discussion of this here and work rather formally when needed.\par
With the above comments in mind, we take two even elements of the $\Z_{2}^{n}$-super translation algebra
\begin{align*}
A =  x^{\mu} P_{\mu} + \theta_{\alpha}^{I} Q_{I}^{\alpha} +\half  z^{JK}Z_{KJ}, & & B =  x'^{\mu} P_{\mu} + {\theta'}_{\alpha}^{I} Q_{I}^{\alpha} +  \half z'^{JK}Z_{KJ},
\end{align*}
were the factor of one half is included to remover the overcounting - we understand $z$ and $Z$ to be graded symmetric. Then using the Campbell--Baker--Hausdorff formula we obtain
\begin{align}\label{eqn:grouplaw}
A\circ B  =&  \left( x^{\mu} + x'^{\mu} + \frac{1}{2} \theta'^{I}_{\beta} \theta^{J}_{\alpha}\delta_{JI}(C\gamma^{\mu})^{\alpha \beta}  \right )P_{\mu}\\
 \nonumber  & + (\theta^{I}_{\alpha} + \theta'^{I}_{\alpha})Q_{I}^{\alpha}\\
  \nonumber  & + \frac{1}{2}\left(z^{JK} + z'^{JK} + {\theta'}_{\beta}^{\:\left(K \right.} {\theta\phantom{'}}_{\alpha}^{\left.J\right )} C^{\alpha \beta} \right)Z_{JK},
\end{align}
where ${\theta'}_{\beta}^{\:\left(K \right.} {\theta\phantom{'}}_{\alpha}^{\left.J\right )}  := \frac{1}{2}\left(  {\theta'}_{\beta}^{K} {\theta}_{\alpha}^{J} + (-1)^{\langle \deg(K) , \deg(J) \rangle } \:{\theta'}_{\beta}^{J} {\theta}_{\alpha}^{K}  \right)$.
\begin{definition}
  The \emph{centrally extended $\Z_{2}^{n}$-Minkowski space-time} is the $\Z_{2}^{n}$-manifold, which we will denote as $\cM_4^{[n]}$ that comes equipped with privileged global  coordinates
$$(x^{\mu}, \theta_{\alpha}^{I}, z^{JK}),$$
where via symmetry we can insist on $J<K$, and the  $\Z_{2}^{n}$-Lie group structure defined via (\ref{eqn:grouplaw}). By construction,  $x^\mu$ is a Lorentz vector, $\theta_\alpha^I$  are Majorana spinors and $z_{IJ}$ are Lorentz scalars.    The \emph{supersymmetry transformations}  on  centrally extended $\Z_{2}^{n}$-Minkowski space-time are given by
\begin{subequations}
\begin{align}
 & x^{\mu} \mapsto x^{\mu} + \frac{1}{4}\epsilon^{I}_{\beta}\theta_{\alpha}^{J}\delta_{JI} (C\gamma^{\mu})^{\alpha \beta}, \label{eqn:supera}\\
 & \theta_{\alpha}^{I} \mapsto \theta_{\alpha}^{I} + \epsilon^{I}_{\alpha}, \label{eqn:superb}\\
 & z^{JK} \mapsto z^{JK} + \frac{1}{2} {\epsilon}_{\beta}^{\:\left(K \right.} {\theta \phantom{'}}_{\alpha}^{\left.J\right )}C^{\alpha \beta},\label{eqn:superc}
\end{align}
\end{subequations}
where $\epsilon_{\alpha}^{I}$ is a Majorana spinor-valued parameter. 
\end{definition}
To be explicit the degrees of the coordinates are as follows:
\begin{align*}
& \deg(x^\mu) = \underline{0}, && (\textnormal{even}) \\
& \deg(\theta^I_\alpha) = \deg(I) \in \Z_{2}^{n}\setminus \{ \underline{0}\}, &&  (\textnormal{odd})\\
& \deg(z^{IJ}) = \deg(I) + \deg(J) \in \Z_{2}^{n}\setminus \{ \underline{0}\}. && (\textnormal{even})
\end{align*}
In particular, we have nilpotent/odd coordinates $\theta$ and  non-nilpotent/even coordinates $z$ yet, in general, we will have non-trivial commutation relations between these coordinates governed by the $\Z_2^n$-degrees. As we have global coordinates we can construct the sheaf theoretical definition of a $\Z_2^n$-manifold directly, in particular $\Z_{2}^{n}$-Minkowski space-time is isomorphic to a $\Z_2^n$-graded Cartesian space. That is,
$$\mathcal{M}_4^{[n]} \cong \big( \R^4 , \: C^\infty_{\R^4}[[\theta, z]]  \big).$$
Naturally, we see that the underlying smooth  manifold is just standard Minkowski space-time. 
\begin{remark}
A $\Z_{2}^{3}$-graded generalisation of the supersymmetry algebra was given by Le Roy \cite{LeRoy:1997} who had applications to the strong interaction in mind. However, he did not present a geometric understanding thereof.
\end{remark}

\subsection{Invariant differential forms}
A direct calculation shows that we have a basis of the left-invariant differential forms given by
\begin{align*}
& e^{\mu} = \rmd x^{\mu} + \frac{1}{4}\rmd\theta_{\beta}^{I}\theta_{\alpha}^{J}\delta_{J I} (C\gamma^{\mu})^{\alpha \beta}, && \psi^{I}_{\alpha} = \rmd \theta_{\alpha}^{I}, & e^{JK} = \rmd z^{JK} - {\theta}_{\beta}^{\:\left(K \right.} { \rmd\theta \phantom{'}}_{\alpha}^{\left.J\right )}C^{\alpha \beta}.
\end{align*}
We consider differential forms on a $\Z_{2}^{n}$-manifolds to be functions on the $\Z_{2}^{n+1}$-manifold built from the tangent bundle, and then appending an additional component to the degree which is zero for the base coordinates and one for the fibre coordinates (see \cite{Covolo:2016}). In particular, we have
\begin{align*}
& \deg(\rmd e^{\mu}) = (\underline{0},1), && \deg(\rmd \psi^{I}_{\alpha}) = (\underline{I},1), & \deg(\rmd z^{JK}) = (\underline{J} + \underline{K},1)\,,.
\end{align*}
where for notational ease we set $ \underline{I} := \deg(I) \in \Z_2^n$, etc. \par 
Note that these conventions are similar to the Deligne--Freed conventions for superdifferential forms \cite{Deligne:1999}. The left-invariant differential forms we can as standard consider as canonical vielbeins. Choosing these vielbeins is no more than a change of basis of the space of one-forms.  In particular, the canonical vielbeiens are not all closed and so we have \emph{torsion}:
\begin{align*}
& \rmd e^{\mu} =  \frac{1}{4}  (C\gamma^{\mu})^{\alpha \beta} \psi^{I}_{\beta} \psi^{J}_{\beta}\delta_{JI}, && \rmd \psi^{I}_{\alpha} = 0, & \rmd e^{JK} = 0.
\end{align*}
Thus, $\Z_{2}^{n}$-Minkowski space-time -- just like the standard case -- is flat, but has a non-vanishing torsion.
\subsection{Superfields}\label{subsec:Superfields}
To fully and correctly understand \emph{superfields} one must employ the categorical notion of the internal Homs - loosely we need to consider maps between $\Z_{2}^{n}$-manifolds that are parametrised by arbitrary $\Z_{2}^{n}$-manifolds. However, formally we can proceed as in the standard super-case (see for example \cite{West:1990,Wess:1992}).\par 
Note that we have to understand maps between $\Z_{2}^{n}$-manifolds when written in terms of local coordinates as formal power series in coordinates of non-zero degree. However, for the case at hand, all the Majorana spinor coordinates are odd and so are nilpotent. The complication -- as with the standard case of extended supersymmetry -- is the fact that power series in $z$ does not truncate. Thus, an unconstrained superfield will be a formal power series, i.e., the theory will have an infinite number of axillary fields. \par 
Expanding any superfield in terms of the formal coordinates we have
$$\Phi^{A}(x, \theta, z) = \phi^A(x) + \theta_{\alpha}^{I}\chi^{\alpha A}_{I}(x) + \textnormal{higher terms},$$
here $A$ is some extra index or collection of indices,  that may be related to spin, or  isospin etc., we will not be more specific here. The  degree of a superfield can be arbitrary, though homogeneous (if not then we can decompose the superfield into homogeneous pieces).  To lowest order the $\Z_{2}^{n}$-supersymmetry transformations are
\begin{align*}
& \delta\phi^{A}(x) \simeq \epsilon_{\alpha}^{I}\chi_{I}^{\alpha A}(x),\\
&\delta \chi_{I}^{\alpha A}(x) \simeq {-} \frac{1}{4}\epsilon_{\beta}^{J}\delta_{JI}(C\gamma^{\mu})^{\beta \alpha}\frac{\partial \phi^{A}}{\partial x^{\mu}\hfill}(x).
\end{align*}
The form of the $\Z_{2}^{n}$-supersymmetry transformations is of course expected by comparison with the standard case of extended supersymmetry.  Note that the transformations mix fields whose  degrees are related by $\deg(I)$, which we have taken to be odd. Thus, we have a  generalisation of standard (quasi-classical) supersymmetry. The similarity here with Green's parastatistics should be noted (see \cite{Green:1953}). In particular, if we take the superfield to be degree zero, i.e., $\deg\big(\Phi \big) = \underline{0}$, then $\phi^A(x)$ is also degree zero and so  could represent a physical boson. Similarly, $\chi_{I}^{\alpha A}(x)$ are then  odd and thus interpreted to be nilpotent Majorana spinor fields. However, in general pairs of such  Majorana spinor fields need \emph{not} anticommute due to the   $\Z_{2}^{n}$-grading.  Thus, the `physical content' of the theory resembles (to lowest order) a system consisting of  bosons and  parafermions (written in terms of their `Green components') together with a `parasupersymmetry'.
\begin{remark}
 To our knowledge, the first work on parasupersymmetry in the context of quantum mechanics was that of Rubakov \& Spiridonov \cite{Rubakov:1988}. The earliest work that we are aware of that discusses a parafermion generalisation of supersymmetry in a field theoretical setting  is Jarvis  \cite{Jarvis:1978}.   
\end{remark}
\begin{remark}
It is known that `reasonable' systems of paraparticles in relativistic quantum field theory  are equivalent to systems of particles with standard statistics that carry an extra quantum number, see for example Araki \cite{Araki:1961}, Doplicher \& Roberts \cite{Doplicher:1990}, and  Dr\"{u}hl, Haag \& Roberts \cite{Druhl:1970}. Generically, given an algebra of physical observables, there will always be some ambiguity in the non-observable field content. Very loosely, the ambiguity in field content allows a redefinition (a  Klein transformation) of the fields such that the redefined fields obey standard statistics. Thus, the general belief is that the  difference between parastatistics and normal statistics is a matter of choice with regards to the field content of the theory under study. In particular, via field redefinitions, we can keep the well known spin-statistics theorem.  However, the known  equivalence of paraparticles and standard particles only holds for theories that satisfy the DHR superselection constraints, which loosely say that all observable events are localised in space and time. These constraints are not satisfied long-range interactions such as those found in electromagnetic theory. Moreover, interacting field theories and their superselection rules are far less well understood than free theories --  this is a current generic problem with algebraic field theory. A good discussion of the equivalence of paraparticles and particles, together with the limitations of what is mathematically established,  can be found in \cite{Baker:2015}.  Thus, there is still some possibility of,  as of yet unseen (c.f. \cite{Greenberg:1965}),  fundamental parabosons and parafermions.  Indeed, it has even been suggested that paraparticles may be associated with dark matter, see \cite{Kitabayashi:2018,Nelson:2016}. Even more remarkable is the suggestion that parastatistics can be used to solve the problem of the cosmological constant, see \cite{Moffat:2005}.
\end{remark}

\subsection{$\Z_{2}^{2}$-Minkowski space-time}\label{subsec:Z2Minkowski}
As a specific and illustrative example let us consider $\cM_4^{[2]}$. This simplifies some of the expressions and resembles $\mathcal{N}=2$ extended superspace closely. However, one has to be careful with the signs involved with $\Z_2^2$-graded commutators.   Because of the symmetry of $Z_{IJ}$ for the $n=2$ case,  we can rewrite the non-trivial part of the $\Z_{2}^{2}$-super translation algebra  as
$$[Q_{I}^{\alpha}, Q_{J}^{\beta}] = \frac{1}{2}\delta_{IJ}(C\gamma^{\mu})^{\alpha \beta}P_{\mu} + \frac{1}{2}C^{\alpha \beta}(\sigma_{1})_{IJ}Z,$$
where
$$\sigma_{1} : = \begin{pmatrix} 0 & 1 \\ 1 & 0 \end{pmatrix}$$ 
is the first Pauli matrix.  Here $I,J \in \{1,2 \}$. Note that when $I=J$  we have an anticommutator in the `ungraded' language. This is of course in parallel with standard  $\mathcal{N}=2$ supersymmetry. When $I \neq J$ we have a commutator and this is in contrast to the standard case.  
\begin{remark}
Interestingly, there seems to be no direct analogue of the Bogomol'nyi-Prasad-Sommerfield (BPS) bound: there is no direct inequality like `$m \geq |z|$', where $m$ is the mass of a particle. For one, such an inequality is hard to understand for formal variables: we have to understand $z$ as being of degree $(1,1)$ and so cannot simply be a real or complex number. Secondly, there is a subtle sign change in the commutator $[Q^{\alpha}_{I},Q^{\beta}_{J}]$ when $ I \neq J$ as compared with the standard case. This implies that we cannot use the standard reasoning to deduce the BPS bound as we do not have a positive definite operator when $ I \neq J$.  BPS states in standard supersymmetric theories  are generically important as they can be determined using non-perturbative methods, they are one of the `gifts' of supersymmetry. However, the situation for $\Z_{2}^{n}$-graded theories is far less clear.
\end{remark}
Using constructions of Subsection \ref{subsec:ZnMinkowski}, we see that $\Z_{2}^{2}$-Minkowski space-time comes equipped with global coordinates
$$(\underbrace{x^{\mu}}_{(0,0)} , \:  \underbrace{\theta_{\alpha}^{1}}_{(0,1)},\:  \underbrace{\theta_{\beta}^{2}}_{(1,0)},\:  \underbrace{z}_{(1,1)}   ).$$
Explicitly the  $\Z_2^2$-commutation rules are:
\begin{align*}
& \theta_\alpha^1 \theta_\beta^2  = {+} \theta_\beta^2 \theta_\alpha^1 , && z \theta_\alpha^I = {- } \theta_\alpha^I z,
\end{align*}
with all other commutation rules being the same as the standard case of $\mathcal{N}=2$ supersymmetry. Naturally,  we still, have $\big(\theta^I_\alpha\big)^2 =0$, and while $z$ picks up  a minus sign when commuted with any $\theta$, it is not nilpotent (it is a formal coordinate).\par 
The supersymmetry transformations \eqref{eqn:supera},\eqref{eqn:superb} and \eqref{eqn:superc} simplify to
\begin{align*}
 & x^{\mu} \mapsto x^{\mu} + \frac{1}{4}\epsilon^{I}_{\beta}\theta_{\alpha}^{J}\delta_{JI} (C\gamma^{\mu})^{\alpha \beta},\\
 & \theta_{\alpha}^{I} \mapsto \theta_{\alpha}^{I} + \epsilon^{I}_{\alpha},\\
 & z \mapsto z + \frac{1}{4} \epsilon^{I}_{\beta} \theta_{\alpha}^{J}(\sigma_{1})_{JI}C^{\alpha \beta}.
\end{align*}
\noindent The left-invariant vector fields of the related group action are easily derived and are given by (using standard notation)
\begin{align*}
& P_{\mu} = \frac{\partial}{\partial x^{\mu}}, \\
& Q_{I}^{\alpha} = \frac{\partial }{\partial \theta_{\alpha}^{I}} + \frac{1}{4}\theta_{\beta}^{J}\delta_{JI}(C\gamma^{\mu})^{\beta \alpha}\frac{\partial}{\partial x^{\mu}} + \frac{1}{4} \theta_{\beta}^{J}(\sigma_{1})_{JI}C^{\beta \alpha}\frac{\partial}{\partial z},\\
& Z = \frac{\partial}{\partial z}.
\end{align*}
As standard, we will call the vector fields $Q$ the \emph{SUSY generators}. We can also directly find the right-invariant vector fields, and doing so leads to the \emph{SUSY covariant derivatives} 
$$D_{I}^{\alpha} = \frac{\partial }{\partial \theta_{\alpha}^{I}} {-} \frac{1}{4}\theta_{\beta}^{J}\delta_{JI}(C\gamma^{\mu})^{\beta \alpha}\frac{\partial}{\partial x^{\mu}} {-} \frac{1}{4} \theta_{\beta}^{J}(\sigma_{1})_{JI}C^{\beta \alpha}\frac{\partial}{\partial z}.$$
The reader can directly check that the SUSY covariant derivatives  graded commute with the SUSY generators, i.e., $[D,Q] =0$. Moreover, a direct computation gives the expected result (as it should as we are using right-invariant vector fields)
\begin{equation}\label{eqn:DD}
[D_{I}^{\alpha}, D_{J}^{\beta}] = {-}\frac{1}{2}\delta_{IJ}(C\gamma^{\mu})^{\alpha \beta}P_{\mu} {-} \frac{1}{2}C^{\alpha \beta}(\sigma_{1})_{IJ}Z =  {-} [Q_{I}^{\alpha}, Q_{J}^{\beta}]\,. 
\end{equation}
Note that 
$$\mathcal{D} :=  \Span \left \{ D_{I}^{\alpha} \right\}\,,$$
is a \emph{distribution}, i.e., a local direct factor of $\Vect(\cM_4^{[2]})$, (see \cite{Covolo:2016c} for details of distributions on $\Z_2^n$-supermanifolds). One can see that we have a distribution by verifying that $\big( \partial_x , D, \partial_z \big)$ for a local basis for the vector fields on   $\cM_4^{[2]}$. Following Manin \cite{Manin:1991}, we refer to this distribution as the \emph{canonical $\Z_2^2$-SUSY structure} on $\cM_4^{[2]}$. \par 
Given $\mathcal{D}$,  (or any distribution in fact) we can construct the short exact sequence of modules
$$0 \longrightarrow \mathcal{D} \stackrel{\iota}{\lhook\joinrel\longrightarrow} \Vect(\cM_4^{[2]})  \stackrel{\varphi}{\longrightarrow}  \mathcal{E},$$
where $\mathcal{E} := \Vect(\cM_4^{[2]}) \setminus \mathcal{D}$. Locally the map $\varphi$ is just dropping the components of the vector field that lie in the distribution.  The  \emph{Frobenius curvature} of $\mathcal{D}$ is defined as
\begin{align*}
R  : \: & \mathcal{D} \times \mathcal{D} \longrightarrow \mathcal{E}\\
 & (X,Y) \mapsto  R(X,Y) := \varphi\big([X,Y] \big).
\end{align*}
The \emph{characteristic distribution} of $\mathcal{D}$, denoted $\mathcal{C}$,  consists of all $X \in \mathcal{D}$ such that $R(X, -) =0$. As  $\mathcal{C}$ consists of only the zero vector,  $\mathcal{D}$ is said to be \emph{maximally non-integrable}. To see this we just need to consider  the commutator
\begin{equation}\label{eqn:XD}
[X, D_J^\beta] = X_\alpha^I [D_I^\alpha , D_J^\beta] \pm D_J^\beta \big( X_\alpha^I\big)D_I^\alpha,
\end{equation}
where the coefficients $X_\alpha^I$ may depend on all the coordinates. The second term in the above is clearly an element of $\mathcal{D}$, and so it dropped when looking at the Frobenius curvature.  The first term, due to \eqref{eqn:DD}, is in $\mathcal{E}$ and so to construct the characteristic distribution we need to examine when this vanishes. This condition reduces to a set of linear equations
\begin{align*}
&(C\gamma^\mu)^{\alpha \beta}X_\beta^I =0, && C^{\alpha \beta}X_\beta^I =0.
\end{align*}
Due to the fact that the Dirac gamma matrices are non-singular, the only solution to the above is $X_\beta^I =0$, as required.

\section{Closing remarks}\label{sec:conclusion}
In this  paper, we have suggested one reasonable (if somewhat ad hoc) way to construct a $\Z_{2}^{n}$-manifold version of extended super-Minkowski space-time. We fixed the dimension of the underlying manifold to be four, though one can consider other dimensions.  We have not attempted to build any models here. In the standard theory, one needs to use constraints to obtain a finite expansion of the superfield. This is usually done in terms of the SUSY covariant derivatives -- mathematically these are the right invariant vector fields of the associated super Lie group. It is expected that one can mimic this for  $\Z_{2}^{n}$-case (for low $n$ anyway). Currently, the obstruction to model building is the lack of a workable integration theory on $\Z_{2}^{n}$-manifolds (see Poncin \cite{Poncin:2016} for progress in this direction). However, even with all the mathematical tools in place, it is not clear what the physical relevance of such theories would be. Moreover, it is fair to say that $\Z_2^n$-graded objects are not so well-known though the theoretical physics community.  \par 
Another interesting avenue for future exploration is the study of $\Z_2^n$-graded versions of extended de Sitter and anti-de Sitter superalgebras and their integration to $\Z_2^n$-Lie groups.  The reader should note the absence of Majorana spinors on de Sitter space-time, though this only means that we need to accept spinor objects along with their charge-conjugates in the constructions. Importantly, there is increasing  observational evidence  that the expansion of the Universe is accelerating and that a positive cosmological constant fits the data well. Thus, having a handle on  de Sitter $\Z_2^n$-supergravity  could shed light on the problem of dark energy.   Of course, we are being  speculative, but this for sure motivates further work.

\section*{Acknowledgements}
The author thanks Norbert Poncin for many discussions about $\Z_2^n$-geometry. A special thank you goes to Steven Duplij for many fruitful discussions.  Part of this work was reported on at the miniworkshop  ``Supergeometry and Applications'', University of Luxembourg,  December 14-15, 2017.

\end{document}